\pdfoutput=1
\documentclass[prd,twocolumn,nofootinbib,preprintnumbers,showpacs]{revtex4-1}

\usepackage{amssymb,amsmath,amsfonts,amsthm,latexsym,nicefrac,bm,bbm,esint,braket,url,epsfig,graphicx,epstopdf,color}

\usepackage{chngcntr}
\usepackage{csquotes}

\begin{document}

\preprint{DESY-16-181}

\title{{\Large Corrections to $n_s$ and $n_t$ from high scale physics}}

\author{Benedict J.\ Broy}

\affiliation{Deutsches Elektronen-Synchrotron DESY,\\ Theory Group, 22603 Hamburg, Germany}

\begin{abstract}
When modelling inflaton fluctuations as a free quantum scalar field, the initial vacuum is conventionally imposed at the infinite past. This is called the Bunch-Davies (BD) vacuum. If however an asymptotically Minkowskian past does not exist, this requires modifications. We derive corrections to the scalar spectral index $n_s$ and the tensor tilt $n_t$ descending from arbitrary mixed states or from explicit non-BD initial conditions. The former may stem from some pre-inflationary background and can redshift away whereas the latter are induced by a timelike hypersurface parametrising a physical cut-off. In both cases, we find that corrections scale in parts or fully as $\mathcal O(\epsilon)$ where $\epsilon$ is the first slow-roll parameter. The precise observational footprint is hence dependent on the model driving inflation. Further, we show how the inflationary consistency relation is altered. We thus provide an analytic handle on possible high scale or pre-inflationary physics.
\end{abstract}

 
\maketitle

\section{Introduction}

Cosmic inflation \cite{Linde:1983gd, Starobinsky:1980te} has been well established as the leading paradigm to describe the physics of the early universe. Besides solving the horizon and flatness problem, inflation is furthermore predictive as it provides a mechanism \cite{Mukhanov:1990me} seeding structure formation which is in astonishing agreement with recent observations \cite{Bennett:2012zja, Ade:2015lrj, Ade:2015tva}. 

Typically, inflation is taken to be realised by a scalar field with a nearly shift-symmetric potential thus mimicking the equation of state of a cosmological constant. The shift symmetry is broken by a minimum in which the field may settle, hence inducing a graceful exit. Inducing the primordial density perturbation then comes naturally once promoting the inflaton field to a quantum operator. Precisely, the inflaton is decomposed into a classical background field evolving as determined by its potential, and a fluctuating part that is described as a massless quantum scalar field. The equation of motion for the free field then follows from the interplay of the perturbed stress-energy tensor of the inflaton field with the perturbed part of the linearised Einstein tensor. It has the form of that of a harmonic oscillator with time dependent mass. When taking the limit of the infinite past, the simple harmonic oscillator is recovered and one may impose a Minkowski vacuum as a boundary condition for the fluctuation; this is called the Bunch-Davies vacuum \cite{Bunch:1978yq}, and is the conventional procedure whenever an expanding spacetime is asymptotically Minkowskian at the infinite past.

However, problems can arise when questioning the accessibility of the infinite past, or more generally, a Minkowskian limit at some past infinity. A prominent criticism of the standard procedure outlined above was coined the trans-Planckian problem \cite{Anderson:2000wx, Niemeyer:2000eh, Brandenberger:2000wr, Martin:2000xs, Hui:2001ce, Niemeyer:2001qe, Kaloper:2002uj, Burgess:2002ub, Brandenberger:2002hs, Danielsson:2002qh, Danielsson:2002kx,  Bergstrom:2002yd, Kaloper:2002cs, Goldstein:2002fc, Easther:2002xe, Chung:2003wn, Kaloper:2003nv, Burgess:2003hw, Alberghi:2003am, Martin:2003kp, Ashoorioon:2005ep, Meerburg:2010rp, Kundu:2011sg, Brandenberger:2012aj, Kundu:2013gha, Aravind:2016bnx, Shukla:2016bnu}, and questioned whether or not scales from below the Planck length $l_P$ could leave a signature in the cosmic microwave background (CMB) when stretched across the (event) horizon during inflation. A true consensus was never reached.

In this paper, we revisit some of the original considerations and argue that whenever the infinite past is not accessible, the BD vacuum may not be imposed. Instead, one has to resort to mixed states as vacua, so called $\alpha$-vacua \cite{Allen:1985ux,Mottola:1984ar,Danielsson:2002mb}, which have recently been rediscovered in \cite{Handley:2016ods}. While these are commonly thought to be ill behaved, they nevertheless provide a possible handle on imposing a vacuum at some finite past. Allowing yet unknown physics to settle the debate about the consistency of such vacua, we derive corrections to the inflationary observables $n_s$ and $n_t$ induced by arbitrary mixed states or non-BD initial conditions.

Considering arbitrary mixed states, we find that the correction to the inflationary indices depends on the spectrum of mixed states; namely on the coefficient $|B_k|$ of the negative frequency contribution to the solution of the equation of motion for a mode $k$. It reads
\begin{equation}\label{1}
\delta n_{s,t}\sim 2\epsilon_V |B_k|^2-\frac{d|B_k|^2}{d\ln k} \thinspace .
\end{equation}
Assuming non-BD initial conditions, one finds oscillatory corrections
\begin{equation}\label{2}
\delta n_{s,t}\sim \epsilon_V\cos\left(\frac{2\Lambda}{H_{infl}}\exp{(\epsilon_V\thinspace N_e)}\right) \thinspace ,
\end{equation}
where $N_e\leq 0$ is the number of remaining e-folds, $H_{infl}$ is the inflationary energy scale and $\Lambda$ is the physical momentum cut-off where the initial vacuum is imposed.

The rest of the paper is structured as follows; we begin with a short review of field operators in time dependent backgrounds and establish our notation. Following a quick discussion of the relevant equation of motion for inflationary fluctuations, we discuss Bunch-Davies and non-BD initial conditions and highlight which scenario requires which choice of vacuum. In the main part, we first derive corrections to $n_s$ and $n_t$ from mixed states and continue to study the explicit example of corrections induced from $\alpha$-vacua. We conclude in section \ref{conclusions}.

\section{Theory}

In this section, we review relevant results and establish our notation. We will introduce scalar fields in time dependent backgrounds as well as their quantisation, equation of motion, and vacuum selection in de Sitter space.

\subsection{Scalar fields in time dependent backgrounds}

We begin by summarising the formalism of \cite{Polarski:1995jg, Lesgourgues:1996jc, Kiefer:1998jk, Bergstrom:2002yd, Danielsson:2002kx,Danielsson:2002qh}. We consider a flat FRW background
\begin{equation}
ds^2=a^2(\tau)\left(d\tau^2-d\bold x^2 \right)\thinspace ,
\end{equation}
where $d\tau=a^{-1}dt$. Inflationary fluctuations are usually modelled as a massless scalar field $\phi$
\begin{equation}\label{scalarFieldPerturbations}
\mathcal L=\frac{1}{2}g^{\mu\nu}\partial_\mu \phi \partial_\nu \phi\thinspace ,
\end{equation}
but the equation of motion will be derived for the rescaled field $f=a(\tau)\phi$. When promoting the rescaled field variable to be a field operator, the variable's conjugate momentum is required in order to impose relevant commutation relations. 
The conjugate momentum of the rescaled field is
\begin{align} \label{rescaledcm}
\pi_{f}&=\frac{\partial\mathcal L}{\partial f'}=f'-\frac{a'}{a}f\thinspace ,
\end{align}
and we impose the following commutation relations
\begin{align}
[\hat f(\tau,\bold x),\hat\pi_f(\tau,\bold y)]&=i\delta^{(3)}(\bold x-\bold y)\thinspace , \\
[\hat f(\tau,\bold x), \hat f(\tau,\bold y)]&=[\hat \pi_f(\tau,\bold x),\hat\pi(\tau,\bold y)]=0\thinspace,
\end{align}
and
\begin{equation}
[\hat f(\tau,{\bold k}),\hat\pi^\dagger_f(\tau,{\bold k}')]=i\delta^{(3)}({\bold k}-{\bold k}')\thinspace.
\end{equation}
The creation and annihilation operators satisfy
\begin{equation}
[\hat a_{\bold k}, \hat a^\dagger_{{\bold k}'}]=\delta^{(3)}({\bold k}-{\bold k}'),\ [\hat a_{\bold k}, \hat a_{{\bold k}'}]=[\hat a^\dagger_{\bold k}, \hat a^\dagger_{{\bold k}'}]=0\thinspace.
\end{equation}
Now, consider a component of the field operator $\hat f$
\begin{equation}
\hat f_{\bold k}(\tau)=f_k(\tau)\hat a_{\bold k} + f^*_k(\tau)\hat a^\dagger_{-{\bold k}}\thinspace , \notag
\end{equation}
subject to the normalisation condition 
\begin{equation}\label{normalisationcondition}
\langle f_k,f^*_k \rangle \equiv i(f^*_k  f'_k- f'^*_k f_k) =1\thinspace .
\end{equation}
Having the operators absorb the time dependence gives
\begin{equation}\label{timedepa}
\hat f_{\bold k}(\tau) = \hat a_{\bold k}(\tau)+\hat a^\dagger_{-{\bold k}}(\tau)\thinspace .
\end{equation}
Likewise, we may write the conjugate momentum as 
\begin{equation}
\hat \pi_{\bold k}(\tau) = -ik\left(\hat a_{\bold k}(\tau)-\hat a^\dagger_{-{\bold k}}(\tau)\right)\thinspace .
\end{equation}
We recast the creation and annihilation operators in terms of their values at some fixed time $\tau_0$ as
\begin{align}
\hat a_{\bold k}(\tau)&=u_k(\tau)\hat a_{\bold k}(\tau_0)+v_k(\tau)\hat a^\dagger_{-{\bold k}}(\tau_0) \thinspace ,\label{squeezed}\\
\hat a^\dagger_{-{\bold k}}(\tau)&=u^*_k(\tau)\hat a^\dagger_{-{\bold k}}(\tau_0)+v^*_k(\tau)\hat a_{{\bold k}}(\tau_0)\thinspace ,
\end{align}
which are essentially Bogolubov transformations and yield the mixing of creation and annihilation operators with time. The commutation relations have to obey condition \eqref{normalisationcondition}, we hence have
\begin{equation}
|u_k(\tau)|^2-|v_k(\tau)|^2=1 \thinspace .
\end{equation}
 Substituting the above into equation \eqref{timedepa} yields
\begin{align}
\hat f_{\bold k}(\tau)&= f_k(\tau)\hat a_{\bold k}(\tau_0)+f^*_k(\tau)\hat a^\dagger_{-{\bold k}}(\tau_0)\thinspace , \label{newField}\\
\hat  \pi_{\bold k}(\tau)&= -i\left(g_k(\tau)\hat a_{\bold k}(\tau_0)-g^*_k(\tau)\hat a^\dagger_{-{\bold k}}(\tau_0) \right)\thinspace ,
\end{align}
with
\begin{align}
f_k(\tau)&= \frac{1}{\sqrt{2k}}\left(u_k(\tau)+v_k^*(\tau)\right)\thinspace , \label{modeEquation}\\
g_k(\tau)&= \sqrt{\frac{k}{{2}}}\left(u_k(\tau)-v_k^*(\tau) \right)\label{modeEquation2}\thinspace , 
\end{align}
and where equation \eqref{modeEquation} is a solution to the field equation of the field $\hat f$ and equation \eqref{modeEquation2} may be obtained from the expression for the conjugate momentum. A vacuum may now be defined at some time $\tau_0$ as
\begin{equation}\label{VACUUM}
\hat a_{\bold k}(\tau_0)|0, \tau_0\rangle=0\thinspace .
\end{equation}
Recalling equation \eqref{squeezed}, we can rewrite the above as
\begin{equation}\label{vacuumcondition}
u_k(\tau_0)\hat a_{\bold k}(\tau_0)|0, \tau_0\rangle+v_k(\tau_0)\underbrace{\hat a^\dagger_{-{\bold k}}(\tau_0)|0, \tau_0\rangle}_{=|1,\tau_0\rangle}=0\thinspace ,
\end{equation}
which immediately sets $v_k(\tau_0)=0$ for the above to realise a vacuum. Thus in order to define a vacuum we need mode functions such that this condition is satisfied at the time the vacuum is imposed. 
Subsequently, we focus on the mode function $f_k$ rather than the operator $\hat f_{\boldsymbol{k}}$.

\subsection{Equation of motion in de Sitter space}

The equation of motion determining the evolution of the mode functions $f_k(\tau)$ for the rescaled field $\hat f=a(\tau)\hat\phi$ is the Mukhanov-Sasaki equation \cite{Mukhanov:1990me} (see \cite{Baumann:2009ds} as a standard review)
\begin{equation}\label{mukhanovsasaki}
f''_k+\left(k^2-{z'' \over z}\right)f_k =0\thinspace ,
\end{equation}
where $z={\varphi}'/H$ with $\varphi$ being the unperturbed background value of the inflaton field. It resembles the equation of motion of a harmonic oscillator with time dependent mass. From equation \eqref{rescaledcm}, we infer
\begin{equation}\label{gmomentum}
g_k=f'_k-\frac{a'}{a} f_k\thinspace .
\end{equation}
In de Sitter space with $a=-1/{H\tau}$, we have
\begin{equation}
\frac{z''}{z}\approx \frac{a''}{a} = \frac{2}{\tau^2}\thinspace .
\end{equation}
Equation \eqref{mukhanovsasaki} may be thus recast as
\begin{equation}
f''_k + \left( k^2-\frac{2}{\tau^2}\right)f_k=0\thinspace .
\end{equation}
The above has the general solution
\begin{equation}\label{generalSolution}
f_k(\tau)=A_k \frac{e^{-ik\tau}}{\sqrt{2k}}\left(1-\frac{i}{k\tau}\right)+B_k \frac{e^{ik\tau}}{\sqrt{2k}}\left(1+\frac{i}{k\tau}\right)\thinspace ,
\end{equation}
and the conjugate momentum \eqref{rescaledcm}
\begin{equation}
g_k(\tau)=A_k\sqrt{\frac{k}{2}}e^{-ik\tau}-B_k\sqrt{\frac{k}{2}}e^{ik\tau}\thinspace .
\end{equation}
As we will recall in the following, the asymptotic past of the spacetime under consideration determines which term of the general solution \eqref{generalSolution} exists; we will show that a non-Minkowskian past creates a non-zero Bogolubov coefficient $B_k$ whereas it is zero otherwise.

\subsection{The Bunch-Davies vacuum}\label{bd}

First, we require spacetime to resemble Minkowski space at early times $\tau\to -\infty$. Equation \eqref{mukhanovsasaki} then reduces to a simple harmonic oscillator. We hence specify the initial condition
\begin{equation}\label{MinkowskiInitialCondition}
\lim_{\tau \to -\infty} f_k(\tau) = \frac{1}{\sqrt{2k}}e^{-ik\tau}\thinspace ,
\end{equation}
for the solutions of \eqref{mukhanovsasaki}. Imposing this initial condition on the general solution \eqref{generalSolution} sets $B_k=0\thinspace, A_k=1$ and hence
\begin{equation}\label{uniquesolution}
f_k(\tau)=\frac{e^{-ik\tau}}{\sqrt{2k}}\left(1-\frac{i}{k\tau}\right)\thinspace ,
\end{equation}
as the mode function for the Bunch-Davies vacuum. From equation \eqref{uniquesolution} and \eqref{gmomentum} we quickly find
\begin{align}
u_k&=\frac{1}{2} e^{-ik\tau}\left(2-\frac{i}{k\tau} \right) \thinspace ,\\
v_k&=\frac{1}{2}e^{ik\tau}\frac{i}{k\tau}\thinspace .\label{VKBD}
\end{align}
Furthermore, considering condition \eqref{VACUUM} yields
\begin{align}
&\hat a_k(\tau_0)|0,\tau_0\rangle=  \notag\\
&u_k(\tau_0)\hat a_{\bold k}(\tau_0)|0, \tau_0\rangle +v_k(\tau_0)\underbrace{\hat a^\dagger_{-{\bold k}}(\tau_0)|0, \tau_0\rangle}_{\neq 0}=0\thinspace,
\end{align}
which is readily satisfied for $\tau_0 \to -\infty$. Thus requiring spacetime to resemble Minkowski space for early times consistently yields an initial condition for the Mukhanov-Sasaki equation and satisfies the vacuum condition \eqref{VACUUM}. Note that a sufficient way to choose \eqref{MinkowskiInitialCondition} was going to the infinite and asymptotically Minkowskian past at $\tau\rightarrow-\infty$, having $k\rightarrow\infty$ was not required. The limit $k\to\infty$ should generally not be taken, as the initial vacuum state ought to be imposed for all modes, i.e.\ for all $k$ and at $\tau\to -\infty$. 

However, scales probed by CMB experiments today correspond to modes whose physical wavelength $a/k$ was much below the Planck length $l_P$ at some point during inflation. As physics below the Planck scale is unknown, it is conceptually unclear whether or not the BD vacuum is the correct choice for sub-Planckian modes. This has been coined the trans-Planckian problem and a true consensus has not yet been reached; especially since alternatives are lacking. In the next subsection, we review a competing proposal, namely $\alpha$-vacua which provide a way to impose a vacuum at finite $\tau$ by effectively reinterpreting a mixed state as a vacuum state.

\subsection{$\alpha$-vacua}

So far, we have considered the standard treatment of inflaton fluctuations in de Sitter space. However, assuming the inflationary phase to be of finite duration, the infinite past might not be accessible and thus imposing the Bunch-Davies vacuum at $\tau\to-\infty$ seems inconsistent. In order to satisfy condition \eqref{VACUUM}, we need $v(\tau_0)=0$, yet equation \eqref{VKBD} only approaches zero for $\tau\to-\infty$. Thus the Bunch-Davies vacuum cannot be imposed at a finite time.\footnote{Physically, the inability to impose the Bunch-Davies vacuum at some finite past time can also be understood as follows; as an expanding background always produces particles, any vacuum can only exist when spacetime is (asymptotically) Minkowskian.} In order to find a vacuum state at a finite past time, recall the general solution \eqref{generalSolution}
\begin{equation}
f_k(\tau)=A_k \frac{e^{-ik\tau}}{\sqrt{2k}}\left(1-\frac{i}{k\tau}\right)+B_k \frac{e^{ik\tau}}{\sqrt{2k}}\left(1+\frac{i}{k\tau}\right)\thinspace , 
\end{equation}
and the conjugate momentum \eqref{rescaledcm}
\begin{equation}
g_k(\tau)=A_k\sqrt{\frac{k}{2}}e^{-ik\tau}-B_k\sqrt{\frac{k}{2}}e^{ik\tau}\thinspace . 
\end{equation}
Now recalling equations \eqref{modeEquation} and \eqref{modeEquation2} we may compare them with the functions above and deduce
\begin{align}
u_k&=\frac{1}{2}\left(A_ke^{-ik\tau}\left(2-\frac{i}{k\tau}\right)+B_ke^{ik\tau}\frac{i}{k\tau} \right), \\
v^*_k&=\frac{1}{2}\left(B_ke^{ik\tau}\left(2+\frac{i}{k\tau} \right)-A_ke^{-ik\tau}\frac{i}{k\tau} \right)\thinspace . \label{complexConjugateofv}
\end{align}
We now seek to fix a vacuum at a finite time $\tau_i$ 
\begin{equation}\label{FixingVacuum}
a_k(\tau_i)|0,\tau_i\rangle=0\thinspace .
\end{equation}
Substituting equation \eqref{squeezed} in the above expression yields
\begin{align}
&\left(u_k(\tau_i)\hat a_k(\tau_i)+v_k(\tau_i)\hat a^\dagger_{-k}(\tau_i)\right) |0\thinspace ,\tau_i\rangle \notag \\
=&\underbrace{u_k(\tau_i)\hat a_k(\tau_i) |0,\tau_i\rangle}_{=0}+\underbrace{v_k(\tau_i)\hat a^\dagger_{-k}(\tau_i) |0,\tau_i\rangle}_{\neq 0} \thinspace .
\end{align}
We need $v(\tau_i)=0$ which is satisfied for\footnote{The normalisation condition \eqref{normalisationcondition} yields $|A_k|^2-|B_k|^2=1$. Combined with the requirement $v(\tau_i)=0$, we find the given relation.}
\begin{equation}\label{constraint2}
B_k=\frac{ie^{-2ik\tau_i}}{2k\tau_i+i}A_k\thinspace .
\end{equation}
We observe that a vacuum at a finite past time can only be imposed when the general solution of the Mukhanov-Sasaki equation is taken as the mode function of the rescaled field $\hat f$. 

Since we consider our starting point being the inability to access some infinite and Minkowskian past, we may recognise that by imposing a vacuum at finite $\tau_i$, we have introduced a physical cut-off described by a timelike hypersurface $\Lambda$ (see \cite{Danielsson:2002kx} for more detail); to be more precise: there exists a physical momentum cut-off $\Lambda$  and mode evolution is assumed to begin once $k=a \Lambda$ (note that this is similar to the horizon crossing condition $k=aH$). Hence, for $a=(-H\tau)^{-1}$, the vacuum is imposed for all $k$ at a $k$-dependent initial time
\begin{equation}\label{ktime}
\tau=-\frac{\Lambda}{Hk}\thinspace .
\end{equation}
Considering the normalisation condition $|A_k|^2-|B_k|^2=1$ and expression \eqref{ktime}, the coefficient $B_k$ becomes
\begin{equation}\label{values}
|B_k|^2=\frac{H^2}{4\Lambda^2}\thinspace .
\end{equation}
These vacua are called $\alpha$-vacua as the parameters $A_k$ and $B_k$ may be reparametrised in terms of just one parameter $\alpha$ \cite{Allen:1985ux, Mottola:1984ar, Danielsson:2002mb, Banks:2002nv}. It is however understood that $\alpha$-vacua are non-thermal \cite{Bousso:2001mw} and seemingly violate locality \cite{Banks:2002nv, Einhorn:2002nu}. Furthermore, back-reaction has to remain under control \cite{Tanaka:2000jw, Starobinsky:2001kn}, which also puts limits on the amount of early universe entanglement. It has been argued that unknown physics might help resolve these outstanding issues \cite{Danielsson:2002mb, Danielsson:2005cc}, but this is speculation as of now. It is important to note that $\alpha$-vacua did not arise in the context of trans-Planckian physics but simply due to considering an inflationary phase with finite duration: BD initial conditions can \emph{only} be imposed if an asymptotically Minkowskian past exists and the non-existence of such a history seems to induce non-BD initial conditions, which however come with theoretical obstacles that remain unresolved to date. 

Nevertheless, we will derive corrections to the inflationary scalar and tensor spectra in terms of the number of remaining e-folds in order to provide an easy handle on these effects from the model-building and phenomenology perspective.

\section{Phenomenology}\label{pheno}

We continue with the main part of this work and derive corrections to $n_s$ and $n_t$. After a quick introduction of scale dependence of spectra, we first consider corrections $\delta n_{s,t}$ induced by mixed states \eqref{generalSolution}. Then, we study the example of an $\alpha$-vacuum, i.e.\ a mixed state treated as a vacuum, in detail and give the explicit form of the corrections to the inflationary indices and to the consistency relation. While the explicit form of corrections induced by some mixed state depends on the coefficients $B_k$ and one generically evaluates the spectra at horizon crossing where the oscillatory terms are frozen, $\alpha$-vacua do cause oscillatory corrections with an amplitude of $\mathcal O(\epsilon_V)$.\footnote{For $\alpha$-vacua, the precise expression for $B_k\rightarrow B_k(k)$ is known, namely expression \eqref{values}.}

We will assume $H<<\Lambda$ throughout, i.e.\ the cut-off scale to be at least one order of magnitude above the inflationary energy scale. This will allow to expand in $H/\Lambda$.

\subsection{Two-point functions and spectra}

The general expression is for the two-point function of inflaton fluctuations is

\begin{align}
P_\phi&=\frac{k^3}{2\pi^2a^2}\langle 0 | f^\dagger_k f_{k'} | 0 \rangle \notag \\
&=\frac{k^3}{2\pi^2a^2}\left|A_k\frac{e^{-ik\tau}}{\sqrt{2k}}\left(1-\frac{i}{k\tau} \right) + B_k\frac{e^{ik\tau}}{\sqrt{2k}}\left(1+\frac{i}{k\tau} \right)\right|^2 \notag \\
&=\underbrace{\left(\frac{H}{2\pi} \right)^2}_{\bar P_\phi}\left[1+|B_k|^2\left(2+\frac{2k\tau_i+i}{i}e^{2ik\tau_i}\right.\right. \notag \\ &\quad\quad\quad\quad\quad\left.\left.+\frac{2k\tau_i-i}{-i}e^{-2ik\tau_i} \right) \right]  \equiv \bar P_\phi +\delta P_\phi\label{ModifiedPSpectrum}\thinspace ,
\end{align}
where we have made use of $|A_k|^2-|B_k|^2=1$. Exchanging the exponentials for trigonometric functions, and focusing on the perturbation, one finds
\begin{align}\label{43}
\delta P_\phi&=|B_k|^2\left[2+4k\tau\sin(2k\tau)+2\cos(2k\tau)\right]\thinspace .
\end{align}
The text book approach \cite{Mukhanov:2007zz} then suggests to average over the oscillatory part.\footnote{Literature \cite{Baumann:2009ds} also suggests to evaluate the spectrum at late times, i.e.\ for $k\tau\to 0$. This would induce a factor of two to expression \eqref{43} which we omit.} One arrives at
\begin{equation}\label{spectrumwithcorrection}
P_\phi=\left(\frac{H}{2\pi}\right)^2\left(1+2|B_k|^2\right)\thinspace .
\end{equation}
Depending on whether or not the wavenumber $k$ to which there exists a non-zero $B_k$ lies within the range of observable modes, the correction may be observed in the CMB. Otherwise, it redshifts away. Different scenarios of pre-inflationary physics may induce different $B_k$ and attention to the details of the assumed scenario is important to examine whether or not corrections may be seen.\footnote{Scenario \cite{Cicoli:2014bja} e.g.\ leads to oscillations from pre-inflationary non-de Sitter backgrounds.}

For the example of a physical cut-off as in \eqref{values}, the averaging over the oscillatory part must not be done due to the assumed mode emergence at $k=a\Lambda$. Instead, we find the leading order correction
\begin{equation}\label{Pcorrections}
\delta P_\phi= \frac{H}{\Lambda}\sin\left(2 \frac{\Lambda}{H} \right)+\mathcal O^{(2)}\left(\frac{H}{\Lambda}\right) \thinspace.
\end{equation}
Recalling the expression for the power spectrum of metric fluctuations at horizon exit
\begin{equation}
\Delta_s^2=\frac{H^2}{\dot\varphi^2}P_{\phi}=\frac{H^2}{\dot\varphi^2}\left(\bar P_\phi+\delta P_\phi\right)\thinspace,
\end{equation}
where $\varphi$ is the inflaton and $\phi$ the scalar field parametrising the inflaton fluctuation, one then arrives at the corrected expressions for scalar and tensor perturbations
\begin{align}
\Delta_s^2&=\frac{1}{8\pi^2}\thinspace\frac{H^2}{\epsilon}\thinspace\left[1+ \frac{H}{\Lambda}\sin\left(\frac{2\Lambda}{H}\right)+\ldots\right]\thinspace ,\label{deltas} \\
\Delta_t^2&=2\left(\frac{H}{\pi}\right)^2\left[1+ \frac{H}{\Lambda}\sin\left(\frac{2\Lambda}{H}\right)+\ldots\right]\thinspace\label{deltat} ,
\end{align}
where we have quoted the result for tensor perturbations without proof.\footnote{Each polarisation of primordial gravitational waves can be described by a massless scalar field, rescaled with the inverse scale factor. We have hence simply quoted the result for tensor perturbations.}
When considering corrections to scalar and tensor spectrum, their ratio $r$ remains unchanged and is still given as
\begin{equation}\label{tensortoscalarratio}
r=\frac{\Delta_t^2}{\Delta_s^2}=16\epsilon\approx 16\epsilon_V\thinspace .
\end{equation}
Measuring $r$ and $n_t$ then may provide a definite answer to the question about the initial conditions of inflationary fluctuations as we will describe in the following.

\subsection{Scale dependence}

Inflation does not resemble perfect de Sitter but has an end, thus $H$ is slowly varying with respect to the scales exiting the horizon. The scale dependence of the inflationary spectra is quantified as
\begin{equation}\label{indicies}
n_s-1=\frac{d\ln\Delta_s^2}{d\ln k}\thinspace,\quad\text{and}\quad n_t=\frac{d\ln\Delta_t^2}{d\ln k}\thinspace .
\end{equation}
A change of variables 
\begin{equation}
\frac{d}{d\ln k}\left(\thinspace\right)=\frac{d}{dN_e}\left(\thinspace\right)\frac{dN_e}{d\ln k}\thinspace ,
\end{equation}
and the results $dN_e/d\ln k\approx 1+\epsilon$ and $d\ln\epsilon/dN_e=2(\epsilon-\eta)$  evaluate expressions \eqref{indicies} with the vanilla spectra 
\begin{align}
\Delta_s^2=\frac{1}{8\pi^2}\frac{H^2}{\epsilon}\thinspace \quad \text{and}\quad \Delta_t^2=2\left(\frac{H}{\pi}\right)^2\thinspace\label{vanilla} ,
\end{align}
to
\begin{equation}
n_s-1=2\eta-4\epsilon\approx 2\eta_V-6\epsilon_V\thinspace,\quad n_t=-2\epsilon\approx -2\epsilon_V\thinspace,
\end{equation}
where the subscript $V$ denotes the potential slow-roll parameters. In the following, we evaluate expressions \eqref{indicies} with the corrected spectra \eqref{spectrumwithcorrection}, \eqref{deltas} and \eqref{deltat}.

\subsection{Corrections from mixed states}

We begin with the derivation of the generic form of $\delta n_{s,t}$ induced by mixed states. First, we recall that
\begin{equation}\label{targets}
n_s-1=\frac{d\ln\Delta_s^2}{d\ln k}=\frac{1}{\Delta_s^2}\frac{d\Delta_s^2}{dN_e}\underbrace{\frac{dN_e}{d\ln k}}_{1+\epsilon}\thinspace ,
\end{equation}
where we will focus on the $d\Delta_s^2/dN_e$ term in the following; we separate
\begin{equation}
\Delta_s^2\rightarrow\Delta_{s,0}^2+\delta\Delta_s^2=\frac{1}{8\pi^2}\frac{H^2}{\epsilon}+ \frac{2}{8\pi^2}\frac{H^2}{\epsilon}|B_k|^2,
\end{equation}
and expand the $1/\Delta_s^2$ factor
\begin{equation}
\frac{1}{\Delta_s^2}\approx\frac{1}{\Delta_{s,0}^2}+\frac{\delta\Delta_s^2}{(\Delta_s^2)^2}\approx\frac{1}{\Delta_{s,0}^2}\left[1+\mathcal O^{(2)}\left(\frac{|B_k|}{H}\right)\right]\thinspace .
\end{equation}
The leading order terms thus are
\begin{equation}\label{leading}
\frac{1}{\Delta_s^2}\frac{d\Delta_s^2}{dN_e}=\frac{1}{\Delta_{s,0}^2} \frac{d(\Delta_{s,0}^2)}{dN_e} + \frac{1}{\Delta_{s,0}^2}\frac{d(\delta\Delta_s^2)}{dN_e}\thinspace ,
\end{equation}
where the first term on the right hand side evaluates to the known $2\eta-4\epsilon$. We hence calculate
\begin{align}\notag
\frac{d(\delta\Delta_s^2)}{dN_e}&=2\frac{d}{dN_e}\left[\frac{H^2}{\epsilon}|B_k|^2\right] \\ \notag
&=2\frac{H^2}{\epsilon}\frac{d}{dN_e}|B_k|^2+2|B_k|^2\frac{d}{dN_e}\left[\frac{H^2}{\epsilon}\right]\\ \notag
&=2\frac{H^2}{\epsilon}\frac{d}{dN_e}|B_k|^2+2\frac{H^2}{\epsilon}|B_k|^2\left(2\frac{d\ln H}{dN_e}+\frac{d\epsilon}{dN_e}\right) \\ 
&=2\frac{H^2}{\epsilon}\left(\frac{d}{dN_e}|B_k|^2-2\epsilon|B_k|^2\right)\thinspace,
\end{align}
where have assumed $d\epsilon/dN_e\sim\mathcal O^{(2)}(\epsilon)$ and omitted this subleading contribution. We thus obtain
\begin{equation}\label{mixedcorns}
\delta n_{s}\sim -2\left(2\epsilon_V |B_k|^2-\frac{d|B_k|^2}{d\ln k} \right) \thinspace .
\end{equation}

Considering tensor fluctuations, we separate $\Delta_t^2=\Delta_{t,0}^2+\delta\Delta_t^2$ and approximate $1/\Delta_t^2\approx 1/\Delta_{t,0}^2$, so that the missing piece to calculate is
\begin{align}\notag
\frac{d(\delta\Delta_t^2)}{dN_e}&=2\frac{d}{dN_e}\left[H^2|B_k|^2\right]\thinspace\\ \notag
&={H^2}\left(\frac{d}{dN_e}|B_k|^2-2\epsilon|B_k|^2\right)\thinspace.
\end{align}
We may now write the corrected expression for $n_t$ as
\begin{equation}\label{mixedcornt}
\delta n_{t}\sim -2\left(2\epsilon_V |B_k|^2-\frac{d|B_k|^2}{d\ln k} \right)\thinspace,
\end{equation}
which is equal to the correction to the scalar spectral index.

Note that a suitable spectrum $|B_k|^2$ may in principle account for low-$\ell$ phenomenology such as power suppression or specific outliers in the data. Further, a correction $\delta n_t$ does also change the consistency relation, which we will demonstrate in the final subsection.

\subsection{Corrections from $\alpha$-vacua}\label{correctionstons}

\begin{figure}[t!]
\centering
\includegraphics[scale=0.6]{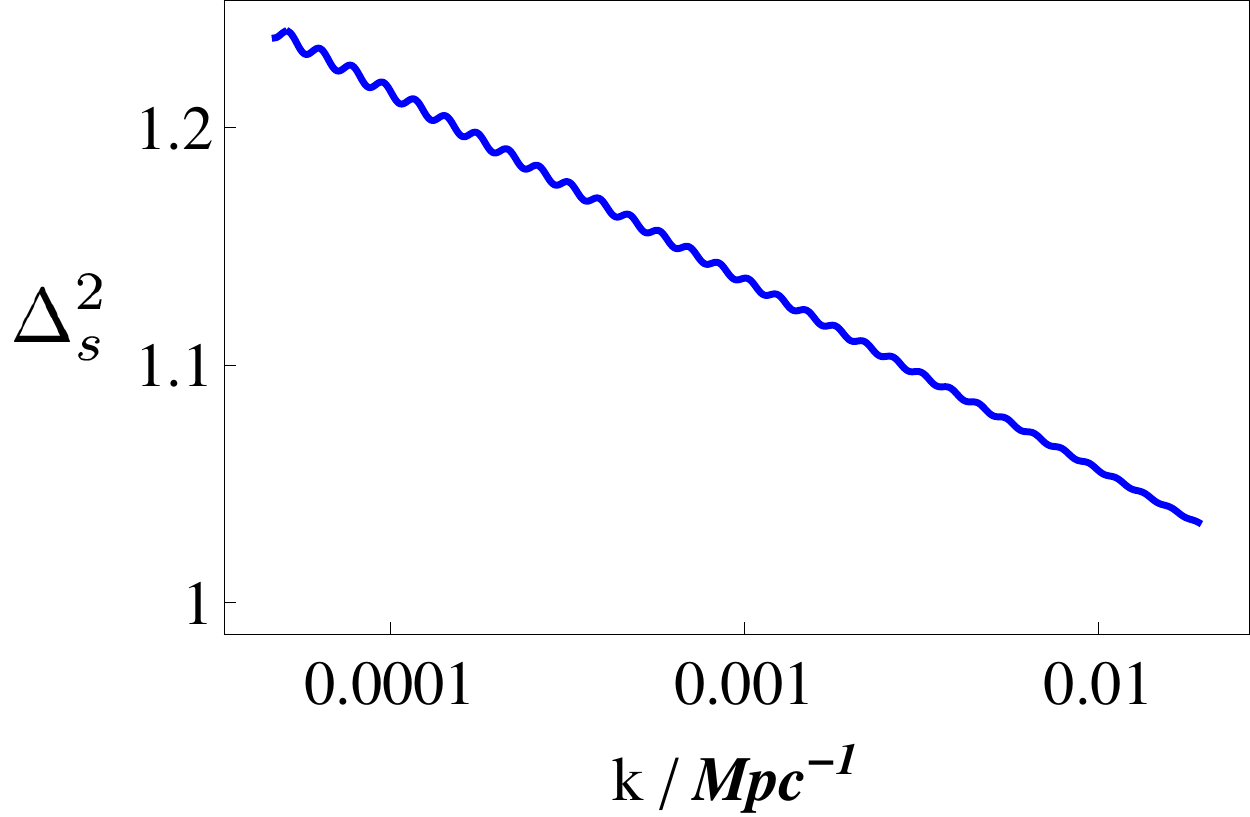}
\caption{\emph{Power spectrum $\Delta_s^2\sim(k/k^*)^{n_s-1}$ between $-62<N_e<-56$ for $V\sim (1-e^{-\kappa\varphi})^2$ with $\kappa=\sqrt{2/3}$, or equivalently $f(R)=R+\alpha R^2$, and pivot scale $k^*=0.05\thinspace Mpc^{-1}$. Here, $\Lambda/H=10^5$. For plateau type models, $\epsilon_V\sim\mathcal O(N_e^{-2})$ whereas $\eta_V\sim\mathcal O(N_e^{-1})$, hence the oscillations are clearly subleading.}}
\label{phystaro}
\end{figure}

\begin{figure}[t!]
\centering
\includegraphics[scale=0.6]{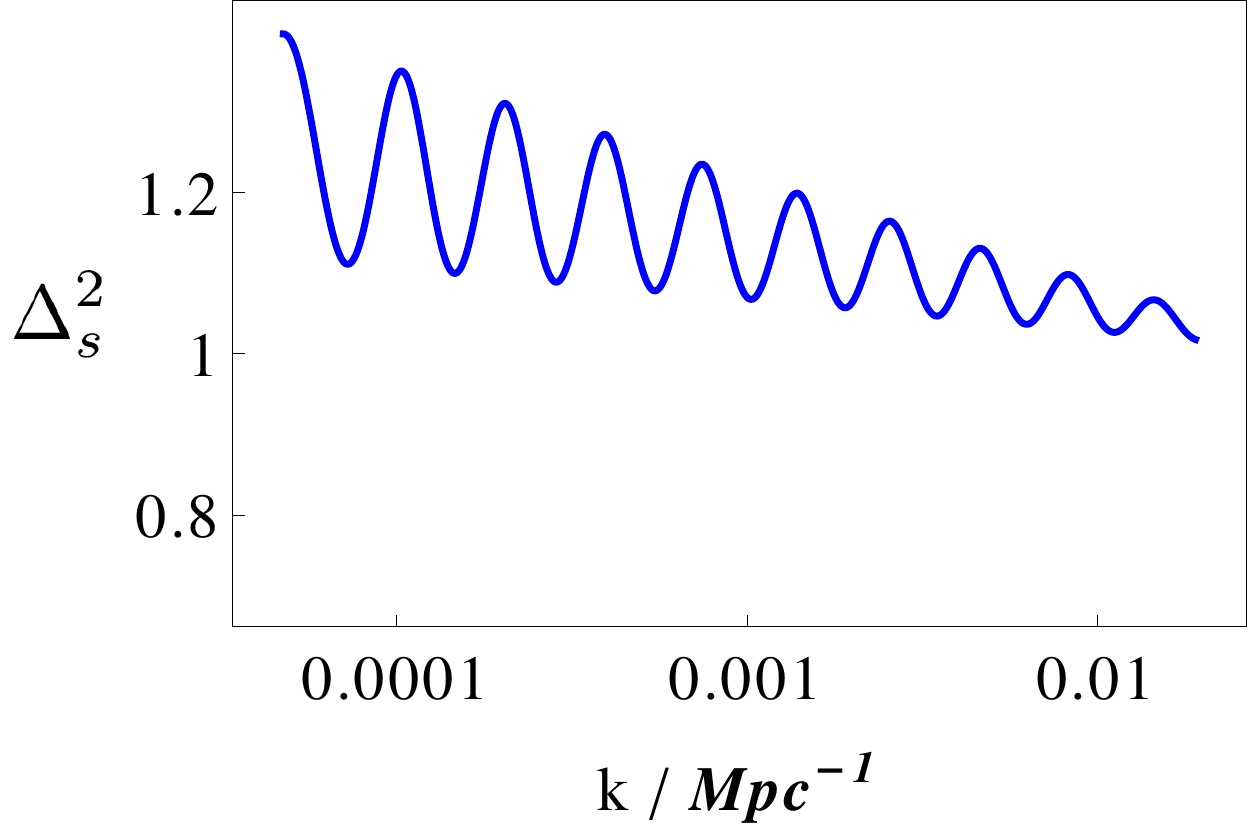}
\caption{\emph{Power spectrum $\Delta_s^2\sim(k/k^*)^{n_s-1}$ between $-62<N_e<-56$ for $V\sim\varphi^2$. Again, $\Lambda/H=10^5$. Since $\eta_V=\epsilon_V$ for $V\sim\varphi^2$, the oscillations are of $\mathcal O(n_s-1)$. Oscillations of this type are surely ruled out by PLANCK.}}
\label{phym2}
\end{figure}

We continue with the calculation of the scalar spectral index with the example of a physical cut-off as an example. Again, we separate \eqref{deltas}
\begin{equation}
\Delta_s^2\rightarrow\Delta_{s,0}^2+\delta\Delta_s^2=\frac{1}{8\pi^2}\frac{H^2}{\epsilon}+ \frac{1}{8\pi^2}\frac{H^3}{\epsilon\Lambda}\sin\left(\frac{2\Lambda}{H}\right)\thinspace,
\end{equation}
and expand the $1/\Delta_s^2$ factor of \eqref{targets} as
\begin{equation}
\frac{1}{\Delta_s^2}\approx\frac{1}{\Delta_{s,0}^2}+\frac{\delta\Delta_s^2}{(\Delta_s^2)^2}\approx\frac{1}{\Delta_{s,0}^2}\left[1+\mathcal O\left(\frac{H}{\Lambda}\right)\right]\thinspace .
\end{equation}
The leading order terms thus are
\begin{equation}\label{leading}
\frac{1}{\Delta_s^2}\frac{d\Delta_s^2}{dN_e}=\frac{1}{\Delta_{s,0}^2} \frac{d(\Delta_{s,0}^2)}{dN_e} + \frac{1}{\Delta_{s,0}^2}\frac{d(\delta\Delta_s^2)}{dN_e}\thinspace ,
\end{equation}
where the first term on the right hand side evaluates to $2\eta-4\epsilon$. We hence calculate
\begin{align}\notag
\frac{d(\delta\Delta_s^2)}{dN_e}&=\frac{d}{dN_e}\left[\frac{1}{8\pi^2}\frac{H^3}{\epsilon\Lambda}\sin\left(\frac{2\Lambda}{H}\right)\right]\\ \notag
&\thinspace\thinspace\thinspace=-\frac{2H}{8\pi^2\epsilon}\cos\left(\frac{2\Lambda}{H}\right)\frac{dH}{dN_e}\\  &\quad+\frac{H^2}{8\pi^2\Lambda\epsilon}\sin\left(\frac{2\Lambda}{H}\right)\left[3\frac{dH}{dN_e}-H\frac{d\ln\epsilon}{dN_e}\right]\thinspace.
\end{align}
Multiplying by $1/\Delta_{s,0}^2$ then has the second term of the right hand side be of $\mathcal O(H/\Lambda)\times \mathcal O(\epsilon,\eta)$. We therefore write
\begin{align}\notag
\frac{1}{\Delta_{s,0}^2}\frac{d(\delta\Delta_s^2)}{dN_e}&=-\frac{2}{H}\cos\left(\frac{2\Lambda}{H}\right)\frac{dH}{dN_e}\\
&=2\thinspace \epsilon\thinspace\cos\left(\frac{2\Lambda}{H}\right)\thinspace .
\end{align}
Having obtained the missing terms of \eqref{leading}, we may now write the result
\begin{align}\notag
n_s-1&=\left[2\eta-4\epsilon+ \epsilon\thinspace \cos\left(\frac{2\Lambda}{H}\right)\right]\left(1+\epsilon\right)\\
&\approx 2\eta_V-6\epsilon_V+ 2\thinspace \epsilon_V\cos\left(\frac{2\Lambda}{H}\right)\thinspace .
\end{align}
We hence find that the scalar spectral index $n_s$ receives oscillatory corrections with an amplitude of $\mathcal O(\epsilon_V)$. Making use of result \eqref{HofN}, we can recast the above in a form more useful to the phenomenologist and write
\begin{equation}\label{nscorrected}
n_s=1+2\eta_V-6\epsilon_V+ 2\thinspace\epsilon_V\cos\left(\frac{2\Lambda}{H_{infl}}\exp{(\epsilon_V\thinspace N_e)}\right)\thinspace ,
\end{equation}
which gives the correction to $n_s$ in terms of the number of e-folds. Figures \ref{phystaro} and \ref{phym2} show how the power spectrum $\Delta_s^2$ receives oscillatory corrections of $\mathcal O(\epsilon_V)$. Thus the amplitude of the observational signature is dependent on the inflationary model realised in nature.



Having obtained the corrected form of $n_s$, we now turn to the tensor tilt $n_t$. Recalling expressions \eqref{deltat} and \eqref{indicies}, we employ an approach similar to the calculation of $n_s$; i.e.\ we separate $\Delta_t^2=\Delta_{t,0}^2+\delta\Delta_t^2$ and approximate $1/\Delta_t^2\approx 1/\Delta_{t,0}^2$, so that the missing piece to calculate is
\begin{align}\notag
\frac{d(\delta\Delta_t^2)}{dN_e}&=\frac{d}{dN_e}\left[\frac{H^3}{\Lambda}\sin\left(\frac{2\Lambda}{H}\right)\right]\\ \notag
&\quad=-2 H\cos\left(\frac{2\Lambda}{H}\right)\frac{dH}{dN_e}\\  &\quad\quad\thinspace+\frac{3H^2}{\Lambda}\sin\left(\frac{2\Lambda}{H}\right)\frac{dH}{dN_e}\thinspace.
\end{align}
Now accounting for the factor $1/\Delta_{t,0}^2$ has the second term of the right hand side be of $\mathcal O(1/\Lambda)$. We thus write
\begin{align}\notag
\frac{1}{\Delta_{t,0}^2}\frac{d(\delta\Delta_t^2)}{dN_e}&=-\frac{2}{H}\cos\left(\frac{2\Lambda}{H}\right)\frac{dH}{dN_e}\\
&=2\thinspace \epsilon\cos\left(\frac{2\Lambda}{H}\right)\thinspace ,
\end{align}
which is equal to the corresponding term of the scalar calculation. As we had separated
\begin{equation}
n_t=\frac{d\Delta_t^2}{d\ln k}=\frac{d\Delta_{t,0}^2}{d\ln k}+\frac{\delta\Delta_t^2}{d\ln k}\thinspace ,
\end{equation}
where the first term of the right hand side is the vanilla result $-2\epsilon$, we may now write the corrected expression for $n_t$ as
\begin{equation}
n_t=-2\left[\epsilon+ \epsilon\cos\left(\frac{2\Lambda}{H}\right)\right]\thinspace .
\end{equation}
Again, assuming slow roll and making use of expression \eqref{HofN}, we write 
\begin{equation}\label{ntcorrected}
n_t=-2\left[\epsilon_V+ \epsilon_V\cos\left(\frac{2\Lambda}{H_{infl}}\exp{(\epsilon_V\thinspace N_e)}\right)\right]\thinspace .
\end{equation}
Importantly, the correction does not change the sign of $n_t$. Note that both \eqref{nscorrected} and \eqref{ntcorrected} do not approach the conventional result for $\Lambda\rightarrow\infty$. However, the starting point of the derivation, i.e.\ \eqref{deltas} and \eqref{deltat} do. 

Further note that the crucial difference to the calculation of corrections from arbitrary mixed states is that the form here is dominated by the oscillatory terms that one had averaged over in the case of mixed states. It is this why the results \eqref{mixedcorns} and \eqref{mixedcornt} do not reduce to \eqref{nscorrected} and \eqref{ntcorrected} once expression \eqref{values} is inserted. Instead, one obtains $\mathcal O(\epsilon_V)\times \mathcal O^{(2)}(H/\Lambda)$ corrections to the oscillatory results.

\subsection{Correction to the consistency relation}

As shown by means of equations \eqref{deltas} and \eqref{deltat}, the tensor-to-scalar ratio $r$ remains unchanged by our considerations. However, as we have just shown that $n_t$ may obtain corrections of order one (conventional result and correction are of $\mathcal O(\epsilon)$), the inflationary consistency relation
\begin{equation}
r=-8n_t
\end{equation}
no longer holds. It is simply verified that, considering \eqref{tensortoscalarratio}, \eqref{ntcorrected}, and \eqref{HofN}, the consistency relation changes to
\begin{equation}
r=8n_t\left[-1+ \cos\left(\frac{2\Lambda}{H_{infl}}\exp{(\epsilon\thinspace N_e)}\right)\right]^{-1}\thinspace .
\end{equation}
Thus measuring the tensor-to-scalar ratio and the tensor tilt may provide a definite answer as to whether or not the duration of inflation is finite \emph{and} a high-scale cut-off of the theory exists, leaving theoretical concerns regarding the use of mixed states as vacua aside for now.

\section{Discussion}\label{conclusions}

In this work, we studied the phenomenology of arbitrary mixed states and non-BD initial conditions. Evaluating expressions for the inflationary observables \eqref{indicies} with the corrected spectra \eqref{spectrumwithcorrection}, \eqref{deltas} and \eqref{deltat}, we found corrections $\delta n_{s,t}$ scaling partly as $\mathcal O(\epsilon_V)$ in the case of arbitrary mixed states and being oscillatory with an amplitude of $\mathcal O(\epsilon_V)$ in the case of non-BD initial conditions. Results \eqref{1} and \eqref{2} thus provide an analytic handle on possible high scale corrections to inflationary observables.

Reflecting on the theoretical motivation for this study, mixed states may be caused by non-slow-roll pre-inflationary backgrounds as described in e.g.\ \cite{Cicoli:2014bja}. There, the duration of inflation is taken to be just the required one, i.e.\ inflation does not last much longer than $|N_{CMB}|$. Concretely, assuming a non-slow-roll and asymptotically Minkowskian background before inflation can in principle induce non-zero Bogolubov coefficients for the inflaton fluctuations (provided the inflaton field already quantum fluctuates before inflation) which hence can induce the comoving excitations described above.

Further, $\alpha$-vacua display several shortcomings. However, it seems that the conventional BD vacuum may only consistently be imposed at the infinite past. Postulating that an asymptotically Minkowskian background at some past infinity is not accessible then renders the standard procedure inconsistent. This could e.g.\ be the case if one assumes the inflationary phase to be without a predecessor, i.e.\ that inflation is indeed the first period in the universe when QFT and GR become applicable in their respective regimes. A consistent formulation of initial conditions, or more generally QFT on curved backgrounds, will surely require a full theory of quantum gravity. 

\begin{acknowledgements}
This work has been supported by the ERC Consolidator Grant STRINGFLATION under the HORIZON 2020 contract no.\ 647995. The author thanks Torsten Bringmann and Alexander Westphal for useful discussions and David Schroeren for helpful comments on the early manuscript.
\end{acknowledgements}


\appendix

\section{From $H$ to $N_e$ to $k$  and back}\label{Relations}

As Inflation has an end and hence does not resemble perfect de Sitter space, the Hubble parameter $H$ is not constant. Variables of interest for phenomenology that parametrise time during inflation are the number of remaining e-folds $N_e$ ($N_e\leq 0$) and the wavenumber $k$ of scales at (event) horizon exit. We now outline how to derive expressions for $H(N_e)$ and $H(k)$ so that we can cast the derived corrections to $n_s$ and $n_t$ in tractable form.

Recall the definition of the Hubble slow-roll parameter 
\begin{equation}
\epsilon=-\frac{\dot H}{ H^2}=-\frac{1}{H}\frac{d\ln H}{dt}=-\frac{d\ln H}{dN_e}\thinspace ,
\end{equation}
where we have made use of $dN_e=dHdt$ for the last equality. Solving for $H$ yields
\begin{align}
\ln \left(\frac{H}{H_{infl}}\right)=-\int_{-|N_{total}|}^{N_e}\epsilon(N_e')dN_e' \thinspace,
\end{align}
where $H_{infl}$ and $N_{total}$  denote the values of the respective quantities at the onset of inflation. Assuming $\epsilon(N_{total})\sim 0$ and $\epsilon$ to be only slowly varying with respect to $N_e$ within the observable range of e-folds, i.e.\
\begin{equation}
\int \epsilon(N_e')dN_e'\approx \epsilon N_e\thinspace,
\end{equation}
we find the expression for $H$ in terms of the number of remaining e-folds to be
\begin{equation}\label{HofN}
H(N_e)=H_{infl}\thinspace \exp{(-\epsilon\thinspace N_e)}\thinspace.
\end{equation}

Now recall the (event) horizon exit condition $k=aH$. Taking the derivative with respect to the wavenumber $k$ and considering the relation $dN_e/d\ln k\approx 1+\epsilon$, one may verify that
\begin{equation}\label{diffEqHofk}
\frac{dH}{dk}=-\epsilon\frac{H}{k}\thinspace .
\end{equation}
This describes the change of $H$ with respect to the scales exiting the horizon. Again, we assume $\epsilon$ to be slowly varying with respect to $k$ for observable scales, i.e.\
\begin{equation}
\int\epsilon(k') k'^{-1}dk'\approx \epsilon \ln k\thinspace .
\end{equation}
Expression \eqref{diffEqHofk} may hence be solved with
\begin{equation}\label{HofK}
H(k)=H_{infl}\left(\frac{k}{k_H} \right)^{-\epsilon}\thinspace .
\end{equation}
Once more, $H_{infl}$ denotes the value of the Hubble parameter at the onset of inflation. The scale $k_H$ is the last scale to exit the horizon, $k$ hence is within $0<k\leq k_H$. As $k$ corresponds to the inverse wavelength, larger $k$ exit the event horizon later. 

The value of $k_H$ can be inferred as follows; wavenumber $k$ and number of remaining e-folds $N_e$ with $N_e\leq 0$ are related as $k=a_{end}H_{infl}\thinspace e^{N_e}$, where $a_{end}$ is the size of the scale factor at the end of inflation ($a_0=1$). Depending on the thermal history of the universe, $a_{end}H_{infl}\approx H_0 e^{\#}$, where $\#\sim 62$. The last scale to exit has $k_H=a_{end}H_{infl}\thinspace e^{0}=a_{end}H_{infl}$. Inserting the expressions for $k_H$ and $k(N_e)$ into \eqref{HofK} then readily recovers \eqref{HofN}.

\bibliography{vacuum}

\end{document}